\documentclass[10pt, conference]{IEEEtran}
\IEEEoverridecommandlockouts

\usepackage{cite}
\usepackage{multicol}

\usepackage{amssymb}

\usepackage{amsmath}
\usepackage{amsfonts}
\usepackage[T1]{fontenc}
\usepackage{algpseudocode}
\usepackage[utf8]{inputenc}
\usepackage[english]{babel}
\usepackage[usenames, dvipsnames]{color}
\usepackage{pifont}
\usepackage{booktabs}
\usepackage{mathtools}
\usepackage{textcomp}
 \usepackage{graphicx}
\usepackage{subfigure}
\usepackage[ruled,vlined,commentsnumbered]{algorithm2e}
\usepackage{amsthm}
\usepackage{setspace}

\usepackage{multirow}

\SetKwInput{KwIn}{Inputs}

\theoremstyle{definition}

\theoremstyle{remark}

\theoremstyle{plain}

\newcommand{\RNum}[1]{\uppercase\expandafter{\romannumeral #1\relax}}

\def\BibTeX{{\rm B\kern-.05em{\sc i\kern-.025em b}\kern-.08em
    T\kern-.1667em\lower.7ex\hbox{E}\kern-.125em}}

\begin{document}
\title{Seeing Through Sparse Signals: CS-Enhanced Sensing With QoS Guarantees for 6G ISAC}

\author{
Yifei Wang$^{*}$, Pradyumna Kumar Bishoyi$^{\dagger}$, and Marina Petrova$^{*}$\\[1pt]
$^{*}$Mobile Communications and Computing (MCC), RWTH Aachen University, Aachen, Germany\\
$^{\dagger}$Department of Electronics Engineering, Indian Institute of Technology Jodhpur, Rajasthan, India\\
Emails: \{yifei.wang, petrova\}@mcc.rwth-aachen.de, pradyumna@iitj.ac.in
}

\maketitle

\begin{abstract}
In integrated sensing and communication (ISAC) systems, communication and sensing share time–frequency resources, creating a fundamental trade-off between communication quality of service (QoS) and quality of sensing (QoSe). Reusing 5G new radio (NR) reference signals (RSs) enables standard-compliant sensing, but their inherent sparsity leads to undersampled channels and strong sidelobes that degrade weak-target detection. Compressed sensing (CS) can mitigate these effects, yet existing methods either overlook communication QoS constraints or rely on reconstruction algorithms that are step-size-sensitive or employ global thresholding, leading to weak-target masking in multi-target scenarios. The multi-target detection problem, therefore, requires a joint design of resource scheduling and CS-based reconstruction. In this work, we investigate the impact of two scheduling schemes, namely QoS-guaranteed scheduling and proportional-fairness (PF)-based scheduling, on the sensing-communication trade-off. To further enhance sensing performance, we employ a 2D max-median complex approximate message passing (2D-mmCAMP) algorithm that performs adaptive row- and column-wise thresholding to suppress sidelobes while preserving weak-target responses in multi-target scenarios. Simulations demonstrate an improvement of $\geq 16\%$ in weak-target detection over baseline reconstruction algorithms while meeting downlink (DL) QoS constraints, validating the effectiveness of the proposed NR-compliant ISAC scheme.\looseness-1

\end{abstract}
\begin{IEEEkeywords}
5G NR, PRS, Compressed Sensing, ISAC, OFDM\looseness-1
\end{IEEEkeywords}

\section{Introduction} \label{sec:Introduction}
Integrated sensing and communication (ISAC) is a key enabler for the next-generation wireless systems, offering seamless integration of high-quality connectivity and sensing capability within a unified framework \cite{ISACfanLiu1,Zhang2022-Survey}. ISAC combines sensing and communication by sharing the same hardware and time-frequency resources, thereby improving efficiency and reducing implementation costs. Orthogonal frequency division multiplexing (OFDM) waveform-based sensing signals have attracted significant attention for ISAC due to their flexibility and compatibility with existing wireless communication standards. In \cite{AthinaEdgeWaveform}, OFDM waveforms are optimized to improve target estimation accuracy. However, such customized waveform designs often deviate from the standardized 5G NR structure, limiting their practical deployment. \looseness-1

In this context, reusing the existing 5G NR reference signals (RSs) for sensing is emerging as a promising and standard-compliant approach \cite{PRSsensing, CSIRSsensingICCworkshop, DMRS+PRS}. Prior works investigated the feasibility of using standard RS, such as synchronization signal (SS), positioning reference signal (PRS), and demodulation reference signal (DMRS), for sensing. Unlike conventional radar systems, which can allocate large sensing resources freely, ISAC must share resource elements (REs) between sensing and communication. Moreover, since sensing and communication inherently pursue different objectives, the joint use of time-frequency resources results in a fundamental trade-off between sensing and communication performance, making resource scheduling a challenging problem in ISAC systems. Therefore, employing sparse RSs for the ISAC system with a low sensing resource budget is a promising approach. Consequently, scheduling RSs for sensing is constrained by the communication Quality of Service (QoS) requirement, resulting in a performance tradeoff between sensing accuracy and communication QoS. Moreover, beyond communication QoS guarantees, future 6G systems must also satisfy Quality of Sensing (QoSe) criteria~\cite{liu2022}. Nevertheless, prior works such as \cite{DMRS+PRS,EsenPRSCAMP} did not investigate the feasibility of scheduling RSs under stringent communication QoS constraints.

\begin{figure*}[t]
	\centering
    \vspace{-10pt}
\end{figure*}

Independent of resource scheduling considerations, the standard 5G NR RSs exhibit sparse and structured patterns in the time-frequency domain, which results in undersampled sensing channel estimates when reused for ISAC.  Taking the QoS into account, the amount of resources available for sensing is further limited. This sparsity in the sensing signal degrades range-Doppler resolution and increases sidelobe levels \cite{YifeiCL}, which often leads to \textit{weak-target masking} in multi-target scenarios. To address this, compressed sensing (CS)-based reconstruction methods are emerging as a popular solution that exploits the inherent sparsity of the delay–Doppler domain and the deterministic RS sampling matrix to recover the missing channel coefficients from under-sampled observations \cite{HighResCSradar,EsenPRSCAMP, BlockSBL_conf}. In \cite{DMRS+PRS}, combined RSs were used for sensing; however, the combination incurs higher overhead, and the resulting degraded communication is not analyzed accordingly. Moreover, the fast iterative shrinkage/thresholding algorithm (FISTA)-based CS reconstruction method adopted in \cite{DMRS+PRS} is sensitive to step-size and suffers from slower convergence. Approximate message passing (AMP) \cite{AMP} and its two-dimensional complex extension (2D-CAMP) \cite{EsenPRSCAMP}, provide faster and more stable convergence for sparse recovery. However, traditional 2D-CAMP uses a global median threshold for denoising, which increases the likelihood of weak-target masking in non-homogeneous multi-target scenarios due to strong target sidelobes from sparse reference signal sampling. \textit{The above observations highlight that sparse NR RS allocation and conventional 2D-CAMP processing cannot maintain reliable detection performance, particularly for weak targets in multi-target environments. Therefore, there is a need for a sensing-centric design that jointly considers resource scheduling and an improved reconstruction algorithm tailored to sparse NR RS patterns.\looseness-1}

To this end, we propose a QoS-guaranteed scheduling scheme for ISAC systems operating under practical NR RS constraints, maximizing the target detectability while satisfying a minimum downlink (DL) communication throughput requirement. To enhance sensing performance under sparse RS and mitigate weak-target masking in non-homogeneous scenarios, we adopt the 2D max-median complex approximate message passing (2D-mmCAMP) algorithm, which suppresses strong sidelobes and improves the weak-target recovery. 
The contributions of this paper are summarized below.
\begin{itemize}
    \item We formulate and solve the resource allocation optimization problem under two scheduling schemes for ISAC systems, namely QoS-guaranteed scheduling (proposed) and proportional-fairness (PF)-based scheduling (baseline). Furthermore, we investigate and evaluate their effect on the trade-off between DL communication throughput and target detection probability. 
    \item We employ a modified 2D-CAMP algorithm, i.e. the 2D-mmCAMP algorithm, that performs adaptive row-/column-wise max-median thresholding to suppress strong sidelobes from sparse NR RS patterns, thereby improving weak-target detectability in non-homogeneous environments at no additional computational cost. 
    \item Extensive simulations demonstrate that the proposed 2D-mmCAMP algorithm reliably resolves weak targets in the presence of strong ones in the Doppler domain, improving the weak target detectability by at least $16\%$ over the 2D-CAMP method. The results further confirm that our scheme maintains the required DL QoS while maximizing sensing performance.

\end{itemize}


\section{System Model}\label{sec:SystemModel}
We consider a scenario where an ISAC-compatible 5G base station (BS) operates at a carrier frequency ($f_c$) of 26 GHz, and performs monostatic sensing and DL communication simultaneously. We assume that the BS operates in full-duplex mode, where the self-interference (SI) is sufficiently low to be cancelled out \cite{liu2022}. We further assume that sensing and communication functionalities are scheduled in a time-frequency-multiplexed manner without interleaving. Let $N_{sym}$ denote the total number of OFDM symbols, $N_{sc}$ denote the total available subcarriers, forming a resource grid with the size of $N_{sc}\times N_{sym}$, where each element in the resource grid is denoted as a RE. Let $N_{RE,sen}$ and $N_{RE, com}$ represent the number of REs allocated for sensing and communication, respectively, such that $N_{RE,sen} + N_{RE,com} = N_{sc}\cdot N_{sym}$. 

\begin{figure}[t]
    \centering
    \includegraphics[width=0.87\linewidth]{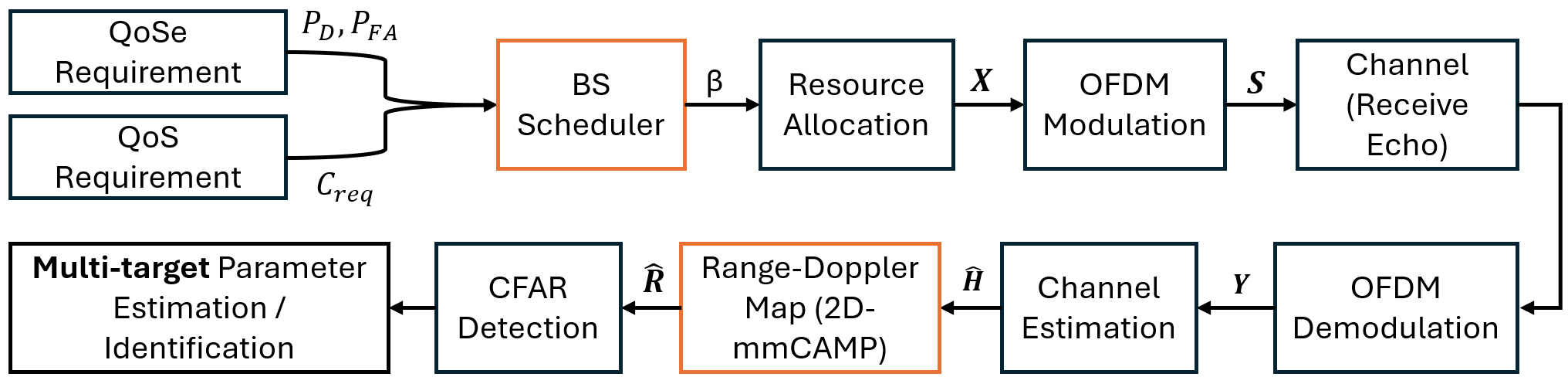}
    \caption{ISAC system processing chain at the base station. The main focus is on the components in orange. }
    \label{fig:System processing chain}
\end{figure}

\subsection{Target Detection QoSe}
Target detection QoSe refers to the process of making binary or multiple decisions about the presence of a specific target based on noisy observations~\cite{Zhang2022-Survey}. It also includes the system's ability to resolve and identify multiple targets. The ISAC-enabled BS detects a target by processing the received echoes reflected from it. The expression for the received echo power is\looseness-1
\begin{equation}
    P_r = \frac{P_{tx}G^2c^2\sigma_{rcs}}{(4\pi)^3f_c^2d^{4}},
\end{equation}
where $P_{tx}$ is the transmit power, $G = 4\pi/\phi^2$ is the gain of the antenna with main beam of width $\phi = [0,2\pi]$, $\sigma_{rcs}$ is the radar cross section (RCS) of the target, $c$ is the speed of light, and $d$ is the distance between the BS and the target. The thermal noise power at the BS receiver is $P_n = k_B T_{therm}FM_{sc}\Delta f$, 
where $k_B$ is the Boltzmann constant, $T_{therm}$ is the standard room temperature (in Kelvin), $M_{SC}$ is the number of subcarriers occupied for that OFDM symbol, $\Delta f$ is the subcarrier spacing (SCS), and $F$ is the noise figure. The corresponding signal-to-noise ratio (SNR) on each OFDM symbol is 
\begin{equation}
    \gamma_{sen} =  \frac{P_r}{P_n} = \frac{P_{tx}G^2c^2\sigma_{rcs}}{(4\pi)^3f_c^2k_B T_{therm}FM_{sc}\Delta fd^{4}}.
\end{equation}
Based on the receiving SNR, the BS makes a decision on whether the target is present or not.

In this work, we consider $P_D$ as the \textit{sensing} key-performance indicator (KPI). Since the sensing SNR depends on the multiplicative gain given by the number of REs utilized for sensing, $P_D$ can be expressed as 
\begin{equation}
    P_D = \mathcal{Q}_1 \left( \sqrt{2\cdot N_{RE,sen}\cdot\gamma_{sen}},\sqrt{-2\cdot ln(P_{FA})}\right),
\end{equation}
where $Q_1(\cdot)$ denotes the first-order Marcum Q-function and $P_{FA}$ denotes the probability of false alarm \cite{skolnik1980introduction} . 

\subsection{Communication QoS}
We consider DL throughput to be the communication KPI of the system. Let $\gamma_{com}$ denote the communication SNR. Considering total $N_{RE, com}$ REs allotted for communication, the total DL communication throughput over the observation period is calculated as
\begin{equation}
    C = \frac{N_{RE,com} \cdot log_2(1 + \gamma_{com})}{N_{sym}\cdot T} \quad bits/s,
\end{equation}
where $T$ is the symbol duration including cyclic prefix (CP).

\begin{figure*}[t]
	\centering
    \vspace{-15pt}
\end{figure*}
\subsection{Sensing Signal Model}
\vspace{-0.1cm}
We assume that the BS uses only 5G NR RSs for sensing. Fig. 1 illustrates the processing chain at the BS performing monostatic sensing, where we focus mainly on scheduling between QoS and QoSe, and range-Doppler (RD) map generation. The scheduler at the ISAC BS first takes the QoSe and QoS requirements, then schedules the corresponding RS(s). The baseband representation of the transmitted sensing signal is 
\begin{equation}
    s(t) = \sum_{n=0}^{N_{sc}-1}\sum_{m=0}^{N_{sym}-1} X_{n,m} e^{j2\pi n \Delta f t} rect\left(\frac{t-mT}{T}\right),
\end{equation}
where $X_{n,m}$ is the QPSK symbol on the $n$-th subcarrier for the $m$-th OFDM symbol, i.e., the $(n,m)$-th element in the symbol matrix $\textbf{X} \in \mathcal{C}^{N_{sc}\times N_{sym}}$, $T = T_{sym} + T_{CP}$ is the total symbol duration, $T_{sym} = 1/ \Delta f$ is the OFDM symbol duration, and $T_{CP}$ is the cyclic prefix duration. Since NR RSs are utilized for sensing, $\textbf{X}$ only has non-zero entries at the certain subcarriers for some OFDM symbols, which can represented by a binary pattern matrix $\textbf{B}\in \{0,1\}^{N_{sc}\times N_{sym}}$. If the $n$-th subcarrier for the $m$-th OFDM symbols is allocated for sensing, $B_{n,m}=1$, and $B_{n,m}=0$, otherwise. 
We consider the line-of-sight (LOS) channel, and the BS receives echos from all $K$ point targets. Therefore, the baseband representation of the received signal at the BS is 
\begin{equation}
    Y_{n,m} = X_{n,m}\sum_{k=1}^K \alpha_k e^{-j2\pi n \Delta f \tau_k}e^{j2\pi mTf_{D,k}} + W_{n,m},
\end{equation}
where $Y_{n,m}$ is the received symbol on the $n$-th subcarrier for the $m$-th OFDM symbol. $\alpha_k$, $\tau_k = \frac{2d_k}{c}$, and $f_{D,k} = \frac{2v_kf_c}{c}$ are the complex channel gain, round-trip delay, and the Doppler shift associated with the $k$-th target, where $c$ is the light speed, $d_k$ and $v_k$ are the distance and the radial velocity of $k$-th target w.r.t the BS, respectively. $W_{n,m} \sim CN(0, \sigma_n^2)$ is the additive white Gaussian noise (AWGN). The BS then performs channel estimation via element-wise division, i.e., 
\begin{equation}
    \hat{H}_{n,m} = \sum_{k=1}^K \alpha_k e^{-j2\pi n \Delta f \tau_k}e^{j2\pi mTf_{D,k}} + Z_{n,m},
\end{equation}

where $\hat{H}_{n,m}$ is the channel estimate for the $n$-th subcarrier of the $m$-th OFDM symbol for $(n,m) \in \{(n,m)| B_{n,m} = 1\}$, and $\hat{H}_{n,m} = 0$, otherwise. $Z_{n,m}$ is the $(n,m)$-th noise symbol in the noise matrix $\textbf{Z}$ after element-wise division. The channel estimate is then utilized for RD map generation, whereas with traditional 2D-FFT, the RD map $\hat{\textbf{R}}$ is obtained by 
\begin{equation}
\small
    \hat{R}(p,q) =  \frac{1}{N_{sc}N_{sym}} \Big|\sum_{n=0}^{N_{per}-1}\sum_{m=0}^{M_{per}-1}\hat{H}_{n,m}e^{j2\pi \frac{pn}{N_{per}}}e^{-j2\pi \frac{qm}{M_{per}}}\Big|^2,
\end{equation}
where $p=0,1,\cdots,N_{per}-1 $, $q=0,1,\cdots,M_{per}-1 $, and $N_{per}$ and $M_{per}$ are the periodogram sizes in the frequency and time domain, respectively. The constant false-alarm rate (CFAR) algorithm is applied on the RD map for target detection and parameter estimation as shown in Fig. \ref{fig:System processing chain}. 

\vspace{-0.15cm}
\section{Compressed Sensing-Augmented Scheduling}\label{sec:Sensing-Centric CandS Scheduling}
Optimizing system resources, such as bandwidth, antenna array size, burst duration, integration time, and transmit power, is essential to achieve high QoSe in ISAC systems \cite{Zhang2022-Survey}. As sensing and communication services share these resources, designing allocation schemes that simultaneously satisfy both QoSe and QoS requirements is inherently challenging. To address this, we investigate the QoSe-QoS trade-off under two distinct scheduling policies: QoS-guaranteed and proportional fairness (PF)-based scheduling. Furthermore, we introduce a CS-based processing algorithm to improve weak target detection performance under limited sensing resources. For the scope of this analysis, we assume that the resource contention between communication and sensing occurs exclusively within the resource-element (RE) domain. We define a time-frequency allocation factor $\beta$,  representing the fraction of the total REs allocated for sensing, i.e., 
\begin{equation}
    \beta = \frac{N_{RE,sen}}{N_{RE}},
\end{equation}
where $N_{RE} = N_{sc} \cdot N_{sym}$ is the total number of REs throughout the available bandwidth and the transmission duration, and $(1-\beta)$ represents the total fraction of resource for communication, which can be further distributed by the scheduler to multiple DL communication UEs based on different QoS requirements. Therefore, the $P_D$ and the DL communication throughput can be rewritten as 
\begin{equation}\label{eq: rewritten PD}
    P_D = \mathcal{Q}_1 \left( \sqrt{2\cdot \beta \cdot N_{RE}\cdot\gamma_{sen}},\sqrt{-2\cdot ln(P_{FA})}\right),
\end{equation}
\begin{equation}\label{eq: rewritten C}
    C = \frac{(1-\beta)\cdot N_{RE} \cdot log_2(1+\gamma_{com})}{N_{sym} \cdot T}.
\end{equation}

\begin{figure*}[t]
	\centering
    \vspace{-10pt}
\end{figure*}
\vspace{-0.21cm}
\subsection{QoS-Guaranteed Scheduling}

For QoS-guaranteed scheduling at the BS, we formulate an optimization problem, where the objective is to maximize the detection probability for the passive targets while satisfying the DL communication throughput, stated in \eqref{eq:OP}
\begin{subequations}\label{eq:OP}
    \begin{align}
    \max_{0\leq \beta \leq 1} \quad \quad & P_D\\
    \textrm{s.t.} \quad \quad & C \geq C_{req},\label{eq: OP CU constraint}
    \end{align}
\end{subequations}
where $C$ is the achieved DL throughput. Substituting Eq. \eqref{eq: rewritten C} into Eq. \eqref{eq: OP CU constraint}, we obtain

\begin{subequations}\label{eq: rewrite CU OP}
    \centering
    \begin{align}
    & \frac{(1-\beta)\cdot N_{RE} \cdot log_2(1+\gamma_{com})}{N_{sym} \cdot T} \geq C_{req},\label{eq: rewrite CU OP 1}\\
    & 0 \leq \beta \leq 1 - \frac{C_{req}\cdot N_{sym}\cdot T}{ N_{RE}\cdot log_2(1+\gamma_{com})}.\label{eq: rewrite CU OP 2}
    \end{align}
\end{subequations}
\begin{figure}[t]
	\centering
    \vspace{-25pt}
\end{figure}
As $P_D$ monotonically increases with $\beta$, the $P_D$ is maximized when $ \small \beta = 1-(C_{req}\cdot N_{sym}\cdot T)/( N_{RE}\cdot log_2(1+\gamma_{com}))$, and the optimal value of $\beta$ is
\begin{equation}\label{eq: beta_opt}
    \beta_{opt} = \left\{\begin{array}{rcl}
           1-\frac{C_{req}\cdot N_{sym}\cdot T}{ N_{RE}\cdot log_2(1+\gamma_{com})}, &C_{req} \leq \frac{N_{RE} \cdot log_2(1+\gamma_{com})}{N_{sym} \cdot T},\\
           0, &\text{otherwise.}
    \end{array}\right.
\end{equation}

\subsection{PF-based Scheduling}
To balance sensing and communication performance without arbitrarily favoring one, proportional fairness (PF)-based scheduling can be applied to maximize the sum of the logarithms of the sensing and communication performance. The formulated optimization problem in this case is
\begin{equation}\label{eq:PF opti}
    \max_{0\leq \beta \leq 1} \quad \quad ln\left(P_D(\beta)\right) + ln\left(\frac{C(\beta)}{C_{max}}\right),
\end{equation}
where $C_{max}$ is the maximum DL communication rate. The logarithm in the PF method prevents one objective from overwhelming the other. Since the formulated optimization problem in \eqref{eq:PF opti} is convex ~\cite{Qfcn}, we can obtain a unique and optimal value of $\beta$, i.e. $\beta_{PF}$.\looseness-1

\begin{algorithm}[hbt!]
\small
\caption{2D-mmCAMP Algorithm}\label{alg:1}
\textbf{Input: }{$\hat{\textbf{H}}$, $\textbf{B}$, $\tau$, $n_{iter}$, $N_{sc}$, $N_{sym}$}\\
\textbf{Output: }{$\tilde{\textbf{D}}_{opt}$}\\
{Set iteration index $l=1$, $\textbf{D}^0 = zeros(N_{sc},N_{sym})$, $\textbf{E}^0 = \hat{\textbf{H}}$\\
\While{$l\leq n_{iter}$}{
  $\tilde{\textbf{D}}^{l} = \mathcal{F}\{\mathcal{F}^{-1}\{\textbf{E}^{l-1}\}\} + \hat{\textbf{D}}^{l-1}$\\
  $\sigma_{r} = median\left(abs(\tilde{\textbf{D}}^l),1\right) / \sqrt{ln(2)}$\\
  $\sigma_{c} = median\left(abs(\tilde{\textbf{D}}^l),2\right) / \sqrt{ln(2)}$\\
  $\sigma = max(\sigma_r, \sigma_c)$\\
  $\hat{\textbf{D}}^l = soft(\tilde{\textbf{D}}^l, \tau\sigma)$\\
  $\textbf{E}^l = \hat{\textbf{H}} - \textbf{B}\odot \mathcal{F}^{-1}\{\mathcal{F}\{\hat{\textbf{D}}^{l}\}\} + \frac{||\hat{\textbf{D}}||_0}{N_{sc}N_{sym}}\textbf{E}^{l-1}$\\
  $l \coloneq l+1$}
  $\tilde{\textbf{D}}_{opt} = \mathcal{F}\{\mathcal{F}^{-1}\{\textbf{E}^{n_{iter}}\}\} + \hat{\textbf{D}}^{n_{iter}}$
  }
\end{algorithm}

For cellular-based ISAC systems with heavy DL communication requirements, the obtained $\beta_{opt}$ or $\beta_{PF}$ is often low. This limits the number of REs allocated for sensing, which results in a decimated channel estimate, thereby reducing the sensing performance. For that, we consider applying compressed sensing-based algorithms for generating the RD map instead of direct usage of periodogram, to improve the detection accuracy. Furthermore, if NR RSs are utilized for sensing, their decimated channel estimate also creates high artifacts and sidelobes along the RD bins for each target in the RD map due to the uniformity and periodicity. The phenomenon does not have an impact on single-target scenarios, however, for non-homogeneous multi-target scenarios, where targets have different RCSs, the weaker targets with less RCSs has a high probability of being masked by the sidelobe of the stronger target if they are closely placed in either Doppler or delay domain. In these cases, if a common median estimate is taken in the CS algorithm for noise reduction, e.g. 2D-CAMP algorithm in \cite{EsenPRSCAMP}, the weak targets are most likely to be missed out. Therefore, we propose to employ a modified 2D-CAMP that takes the concept of the CFAR-AMP in \cite{KnillCAMP}, namely 2D max-median CAMP (2D-mmCAMP), to have adaptive noise floor estimation for each RD bin to mitigate the weak-target masking caused by strong targets nearby.

\begin{figure}[t]
	\centering
    \vspace{-10pt}
\end{figure}
 \begin{figure}
    \centering
    \includegraphics[width=0.97\linewidth]{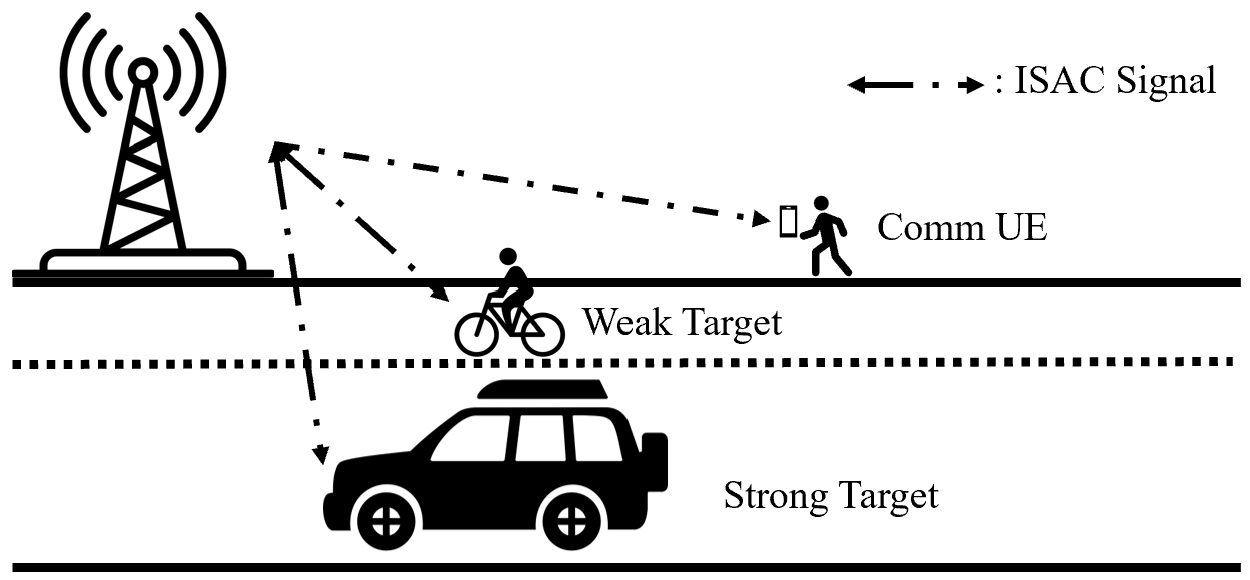}
    \caption{Simulation scenario (BS treated as origin during simulation).  }
    \label{fig:simulation scenario}
\end{figure}

\begin{figure*}[t]
	\centering
    \vspace{-15pt}
\end{figure*}

\begin{figure}[t]
    \centering
    \subfigure[$\gamma_{com}$ vs. $P_D$ and $C$.\label{fig:commSNR vs. beta}]{%
            \includegraphics[width=0.493\linewidth]{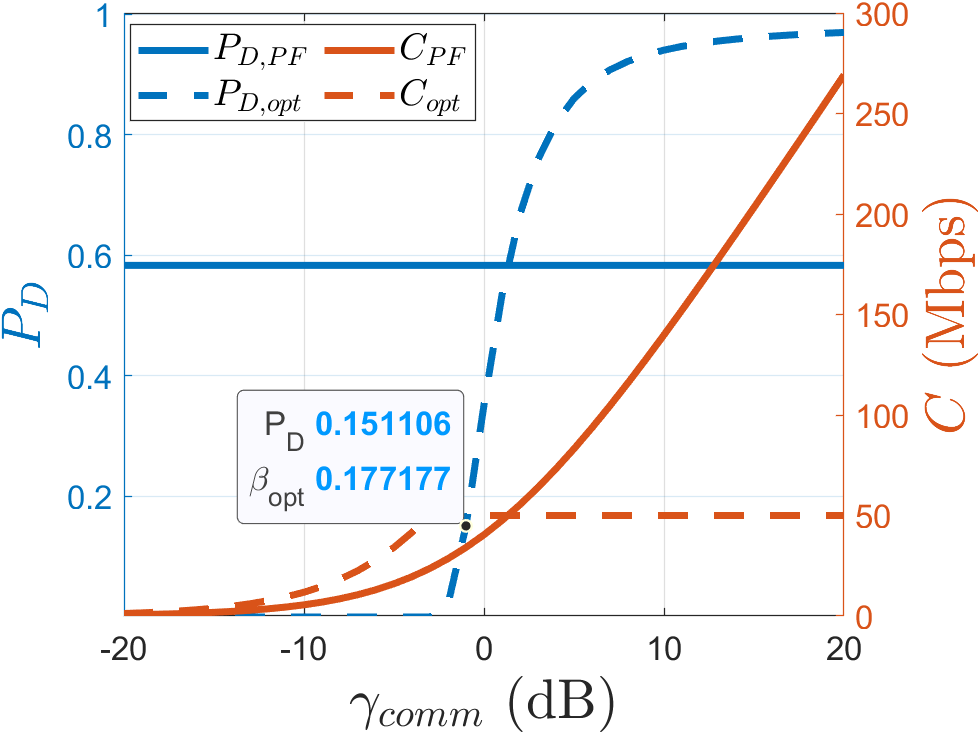}}
    \subfigure[$P_D$ and $C$ for different $C_{req}$.\label{fig:Bar plot}]{%
        \includegraphics[width=0.493\linewidth]{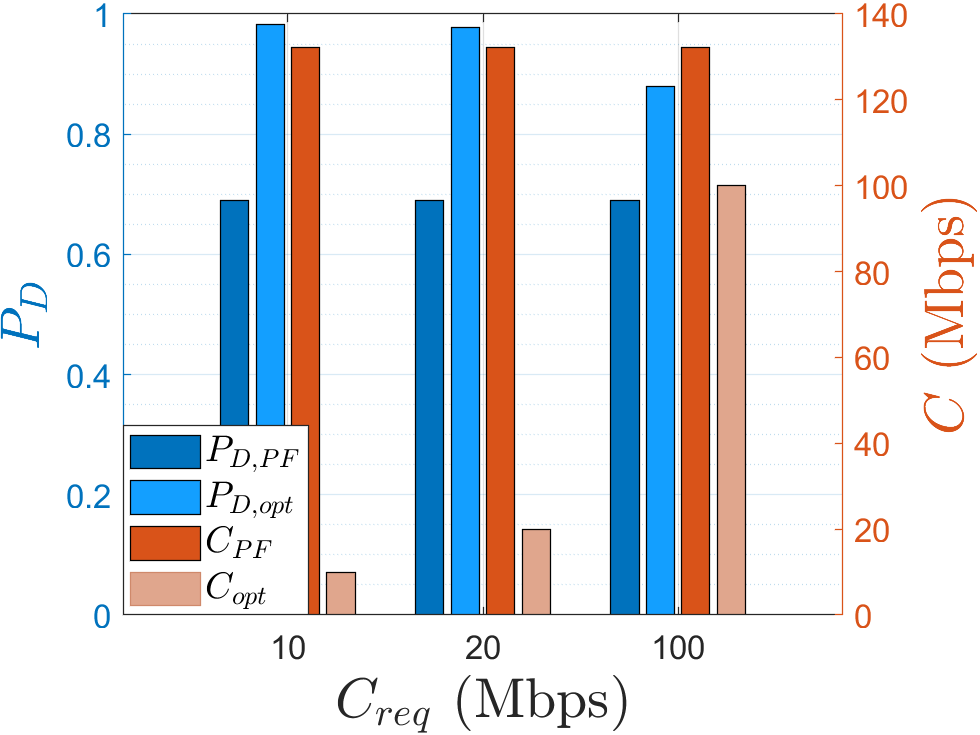}}
    \caption{Tradeoff between DL communication throughput $C$, and probability of detection $P_D$, for QoS-guaranteed and PF-based scheduling for (a) different SNR values $\gamma_{com}$ and (b) different $C_{req}$.}
      \label{fig:Tradeoff between sensing communication performance} 
\end{figure}

\begin{figure}[!t]
    \centering
    \subfigure[PRS6.\label{fig: PRS6 target spacing}]{%
            \includegraphics[width=0.48\linewidth]{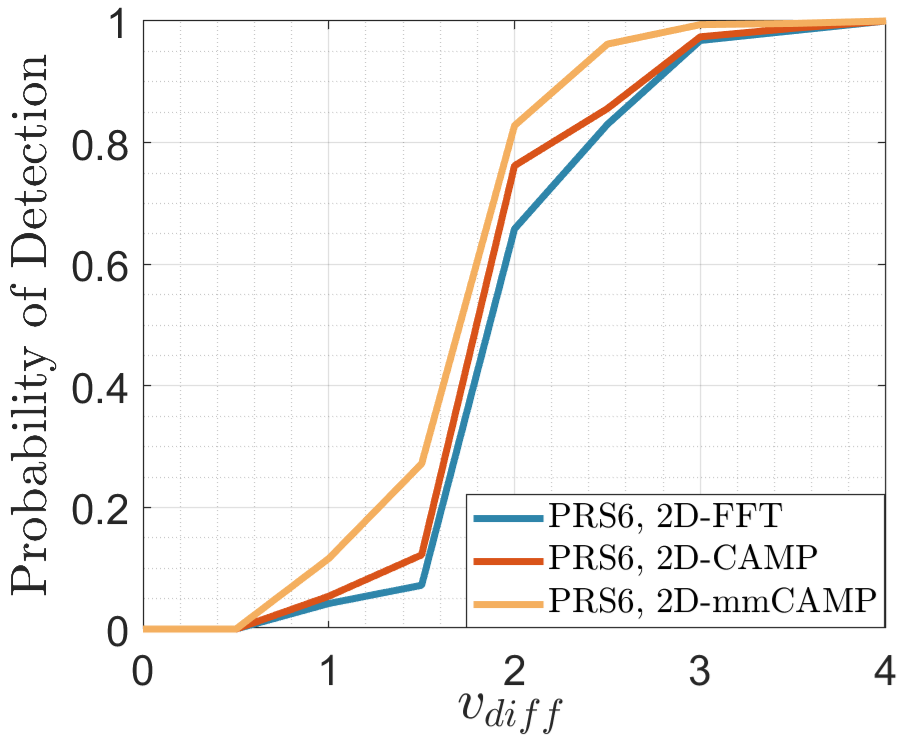}}
      \subfigure[DMRS3.\label{fig: DMRS3 target spacing}]{%
        \includegraphics[width=0.48\linewidth]{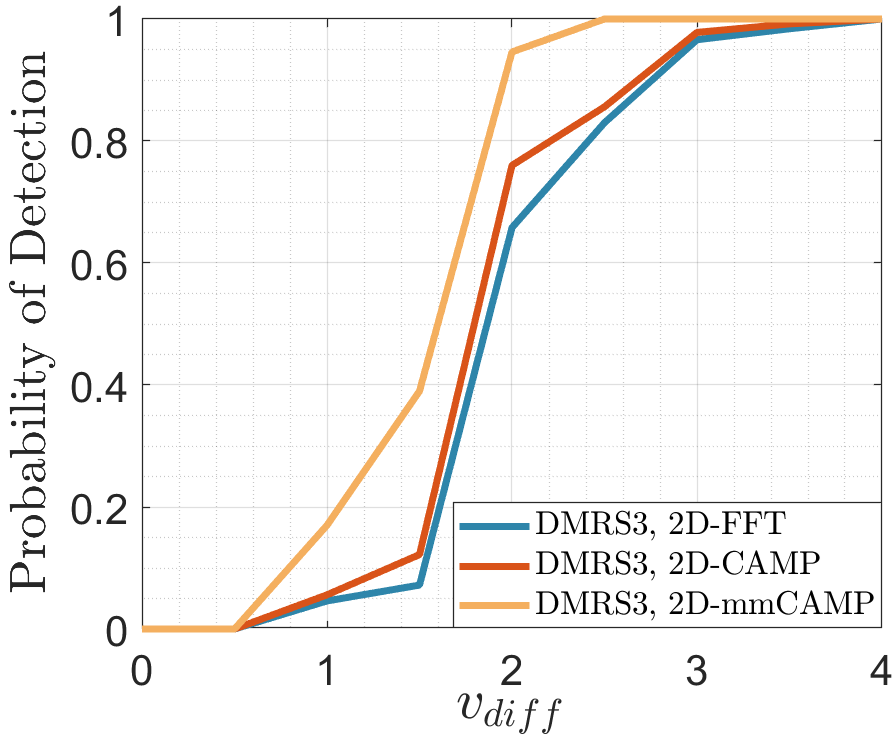}}
      \caption{$P_D$ vs. velocity spacings between targets $v_{diff}$ .}
      \label{fig: PD vs target spacing} 
\end{figure}


\begin{table}[htbp]
\small
\caption{Simulation parameters.}
\begin{center}
\begin{tabular}{|c|c|}
\hline
\textbf{Parameter} & \textbf{Value}\\
\hline
$f_c$ & $26$ GHz \\
\hline
Subcarrier spacing & $120$ kHz\\
\hline
Nr. OFDM symbols & $336$ \\
\hline
Nr. subcarriers & $768$ \\
\hline
EIRP & $43$ dBm \\
\hline
Noise figure & $10$ dB\\
\hline
Maximum target velocity & $20$ $m/s$ ($72$ $km/h$) \\
\hline
Maximum target distance & $104$ $m$ \\
\hline
\end{tabular}
\label{tab:Simulation parameters}
\end{center}
\end{table}

\subsection{2D Max-Median CAMP (2D-mmCAMP)}
The 2D-mmCAMP, summarized in Algorithm \ref{alg:1}, is designed to be more compatible to multi-target scenarios by having adaptive row-/ column-wise thresholding, with no increase in total computational complexity. 

The 2D-mmCAMP algorithm takes the sparse channel estimate $\hat{\textbf{H}}$, the sparse binary pattern matrix $\textbf{B}$, threshold parameter $\tau$, the maximum number of iterations $n_{iter}$, number of subcarriers $N_{sc}$, and number of OFDM symbols $N_{sym}$ as the inputs. Within each iteration $l$, the noisy estimate of the delay-Doppler (DD) matrix $\tilde{\textbf{D}}$ is first calculated. Unlike 2D-CAMP, where a common median is selected taking all elements in $\tilde{\textbf{D}}$, the 2D-mmCAMP calculates the median of each row and each column, stated as $\sigma_r$ and $\sigma_c$, respectively. Furthermore, the maximum value between the median of the row and the median of the column in which the reference cell is located is determined and selected as the denoising threshold parameter for that cell, which is stated as $\sigma$ in the algorithm. The denoised DD matrix $\hat{\textbf{D}}$ is then obtained by performing element-wise soft-thresholding, and the residual $\textbf{E}$ is calculated and utilized in the $l+1$-th iteration of the algorithm. The computational complexity of 2D-mmCAMP is given by $\mathcal{O}\left(n_{iter}N_{sc}N_{sym}\log(N_{sc}N_{sym})\right)$, which is the same as for 2D-CAMP. \looseness-1

\section{Performance Evaluation}\label{sec:PerformanceEvaluation}
In this section, we first investigate the theoretical tradeoff between communication and sensing. Then, evaluate and compare the performance of using single-type RS for sensing with different processing methods through extensive simulations. 

The simulation parameters are shown in Table \ref{tab:Simulation parameters}. Furthermore, we consider the LOS channel and MATLAB ray-tracing is used to model the mmWave propagation environment. The simulation scenario is presented in Fig. \ref{fig:simulation scenario}.

\subsection{Communication and Sensing Tradeoff Evaluation}
Fig. \ref{fig:commSNR vs. beta} illustrates the trade-off between the detection probability and the DL communication rate for varying $\gamma_{com}$ at a fixed sensing SNR. For QoS-guaranteed scheduling, once the required rate $C_{req}=50$ Mbps is met (at $\gamma_{com} = -3$ dB), the scheduler prioritizes sensing by fixing the DL communication rate $C_{opt}$, and increases the amount of resource allocated to sensing. For values of $\gamma_{com} \leq 2$ dB, QoS-guaranteed scheduling achieves higher DL throughput than PF-based scheduling, but at the cost of degraded sensing performance. For example, at a communication SNR $\gamma_{com}=-2$ dB, $\beta_{opt}\approx 1/6$, that yields a $P_{D,opt}=0.15$, which is insufficient for most applications. Therefore, efficient post-processing methods, such as 2D-mmCAMP, is required to improve the target detection probability. For PF-based scheduling, a fixed $\beta_{PF} = 0.53$ balances sensing and communication, resulting in a constant detection probability of $P_{D,PF}=0.58$. The communication rate $C_{PF}$ in result scales with $\gamma_{com}$, but falls below $C_{req}$ in low communication SNR regimes due to the lack of QoS guarantees.\looseness-1

\begin{figure}[!ht]
    \centering
    \subfigure[PRS6.\label{fig: PRS6 pd vs RCS of weak}]{%
            \includegraphics[width=0.48\linewidth]{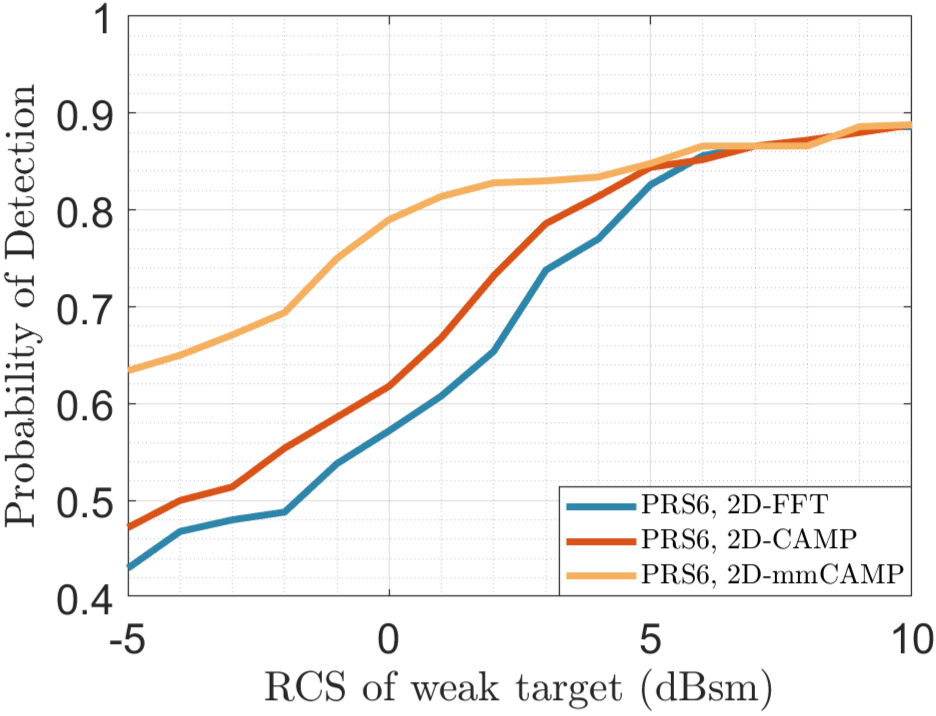}}
      \subfigure[DMRS3.\label{fig: DMRS3 pd vs RCS of weak}]{%
        \includegraphics[width=0.48\linewidth]{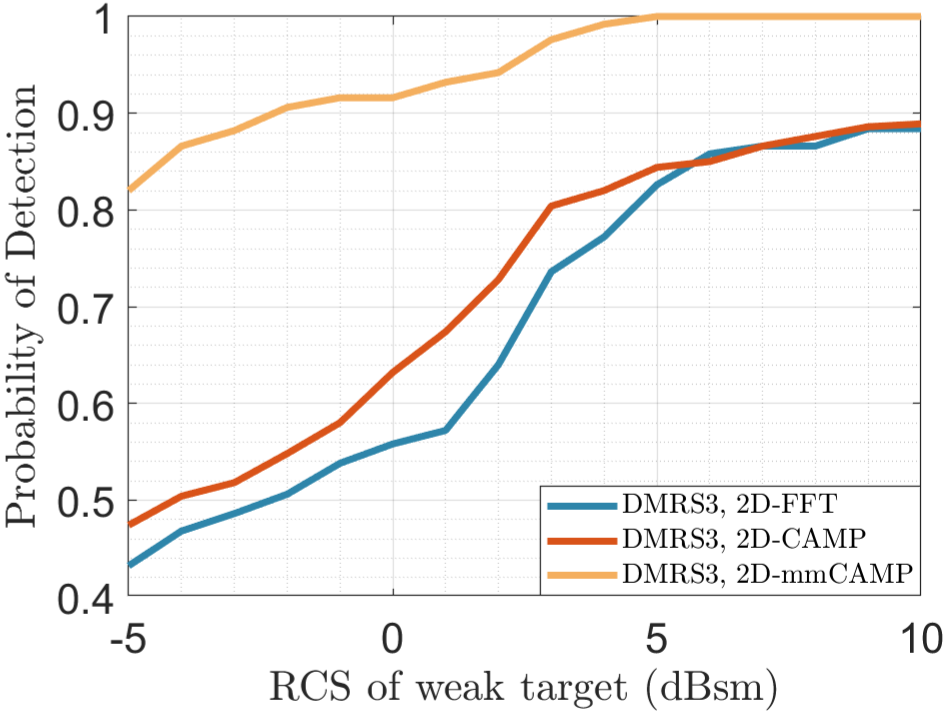}}
      \caption{$P_D$ vs. RCS of the weak target.}
      \label{fig: PD vs RCS of weak} 
\end{figure}

Fig. \ref{fig:Bar plot} shows the $P_D$ and the DL communication rate $C$ for different $C_{req}$.  It can be seen that with a fixed sensing SNR and a good communication channel quality, $P_{D,opt}$ remains around $0.98$ for relatively low $C_{req}$, and for high $C_{req}=100$ Mbps, the $P_{D,opt}$ decreases to $0.88$. At the same time, $P_{D,PF}=0.69$ for all $C_{req}$ as it is only relevant to the sensing SNR. Therefore, if the QoS of $C_{req}$ is satisfied, the QoS-guaranteed scheduler is more flexible and can significantly improve the sensing performance compared to PF-based scheduling.

\subsection{Multi-target Detectability with CS}

In this subsection, we investigate the impact of different post-processing methods on multi-target detection of the ISAC system under low values of $\beta_{opt}$, e.g. for $1/6$ and $1/7$. To ensure compliance with the 5G NR framework, we use the positioning reference signal with a comb size of 6 (PRS6) when $\beta_{opt}=1/6$ and the demodulation reference signal-additional-3 (DMRS3) when $\beta_{opt}=1/7$. We simulate a scenario where two targets are at the same distance, but have different radial velocities and RCSs. The target with a lower RCS is stated as the weak target, and the target with a larger RCS as the strong target. The performance of 2D-mmCAMP is then benchmarked against two post-processing baselines, namely, the periodogram method and 2D-CAMP. 
\begin{figure*}[t]
\vspace{3pt}
    \centering
    \includegraphics[width=0.9\linewidth]{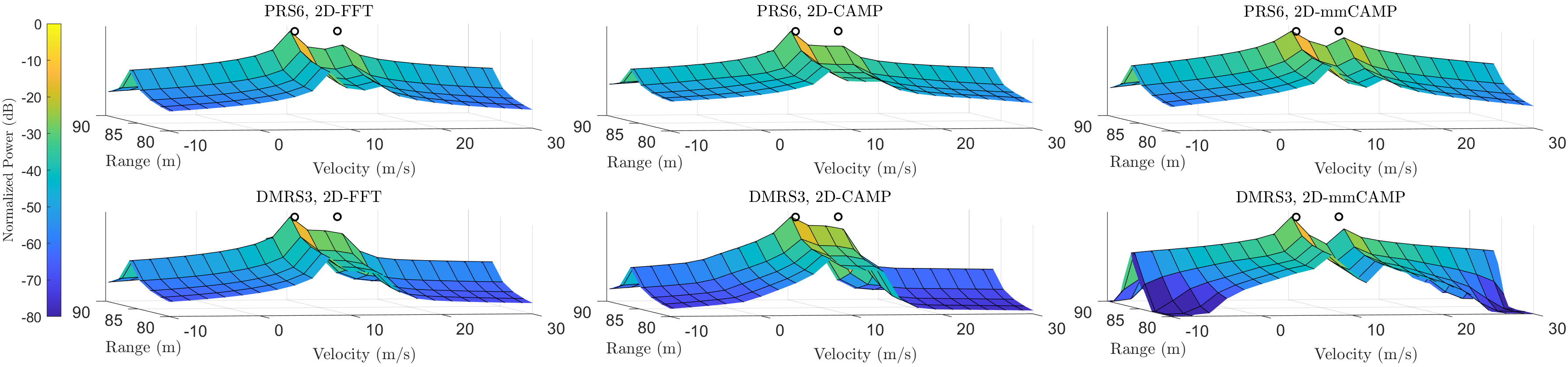}
    \caption{3D heatmaps of RS-based multi-target scenario obtained via 2D-FFT, 2D-CAMP and 2D-mmCAMP.}
    \label{fig:STWPR}
\end{figure*}

\subsubsection{Effect of Doppler Separation}
We first investigate how the spacing between the two targets in the Doppler domain affects the detectability of the weaker target. The velocity resolution is represented by $\Delta v = \frac{c}{2f_cTN_{sym}}$. The RCS of the weak target is fixed at $5$ dBsm, and the strong target RCS at $20$ dBsm. To analyze the masking effect of the strong target to the weak target, we vary the difference in velocity between the two targets, which is quantized by $v_{diff} = \frac{|v_1 - v_2|}{\Delta v}$, and plot the corresponding $P_D$ for the weak target in Fig. \ref{fig: PD vs target spacing}. For both PRS6 and DMRS3, 2D-mmCAMP consistently yields higher $P_D$ than 2D-FFT and 2D-CAMP, particularly when $v_{diff}$ is between 2 and 3. For instance, at $v_{diff}=2.5$, 2D-mmCAMP provides approximately a $10\%$ gain over 2D-CAMP and 2D-FFT for PRS6, and up to $15\%$ improvement for DMRS3. When $v_{diff} >3$, the masking effect from the strong target becomes negligible due to sufficient Doppler separation, and the performance of the three methods converges.

\subsubsection{Low-RCS Target Detectability}
We further investigate the detectability of a low-RCS target when it is closely spaced in the Doppler domain from a strong target. We considered the velocity separation of the two targets to be $v_{diff} = 2.5$. The RCS of the strong target is fixed to be $20$ dBsm, and we change the RCS of the weak target in the range of $-5$ to $10$ dBsm and plot the $P_D$ of the weak target in Fig. \ref{fig: PD vs RCS of weak}. For PRS6 in Fig. \ref{fig: PRS6 pd vs RCS of weak}, it can be seen that 2D-CAMP has a improved $P_D$ compared with 2D-FFT around $4\%$ for RCS of the target below $5$ dBsm, whereas 2D-mmCAMP significantly improves $P_D$ up to $15\%$ compared with 2D-CAMP. For example, when the RCS of the weaker target is $0$ dBsm, its $P_D$ with 2D-mmCAMP is around $0.79$ whereas for the other two methods, the $P_D$ is below $0.65$. For DMRS3, with 2D-mmCAMP, as is shown in Fig. \ref{fig: DMRS3 pd vs RCS of weak}, the $P_D$ for the weaker target reaches $1$ when the RCS of the weak target reaches $5$ dBsm. Whereas for 2D-FFT and 2D-CAMP, the $P_D$ of the weak target only reached 0.85. For low RCS regime, e.g. RCS of the weak target is less than $0$ dBsm, 2D-mmCAMP can bring an increase of $P_D$ up to $34\%$ compared with 2D-CAMP, and up to $38\%$ compared with 2D-FFT.

Fig. \ref{fig:STWPR} further compares the RD heat maps for a two-target scenario obtained with PRS6 and DMRS3 processed by different methods. The RCSs of the targets are set to be $20$ dBsm and $7$ dBsm, respectively, and the velocity separation between the two targets is $v_{diff} = 2.5$. Both targets have the same distance of $86$ m to the BS. The x-axis represents velocity, the y-axis represents range, and the z-axis (color) in the heat maps represents normalized power in dB. The true target speed and range are represented by the black circles, where the velocity for the strong target is $8$ m/s, and $12.7$ m/s for the weak target. For both RSs, it can be seen that although 2D-CAMP can provide gain to the weak target peak, the sidelobe from the strong target is also amplified, decreasing the probability of detection for the weak target when CFAR is applied. However, with 2D-mmCAMP, the weak target peak is more distinct than 2D-FFT and 2D-CAMP because of the adaptive row-/ column-wise selection of the denoising threshold, which improves the detection probability.

\section{Conclusions}
In this work, we investigated resource sharing between sensing and communication in 5G NR-compliant ISAC systems under two types of scheduling algorithms, namely QoS-guaranteed scheduling and PF-based scheduling. For the QoS-guaranteed scheduler, we analytically derived a closed-form expression for optimal RE allocation. Conversely, the unique optimal allocation for the PF-based scheduler was obtained numerically. Under limited sensing resources, we employed the 2D-mmCAMP compressed sensing algorithm to achieve high-resolution range-Doppler image reconstruction and effectively mitigate weak-target masking caused by strong-target sidelobes in non-homogeneous scenarios. Our simulation results confirm that 2D-mmCAMP significantly improves multi-target detection performance over the periodogram method and 2D-CAMP. 
In future work, we will incorporate multi-path propagation models and evaluate different scheduling schemes for ISAC systems.
\vspace{-0.1cm}
\section*{Acknowledgment}
This work was funded by the Federal Ministry of Research,
Technology and Space Germany within the project “SENSa-
tion” under grant 16KIS2529.

\bibliographystyle{IEEEtran}
\bibliography{jsac_ISAC.bib}

\end{document}